\documentclass[doublecol]{epl2} 
\usepackage{amsmath}
\usepackage{amssymb}
\usepackage{amsfonts}
\usepackage{graphicx}
\usepackage{amsthm}
\usepackage{multirow}
\DeclareMathOperator{\Tr}{Tr}

\newcommand{\Ex}{{\rm Ex}}

\newcommand{\ket}[1]{|{#1}\rangle}

\newcommand{\bra}[1]{\langle{#1}|}
\newcommand{\bkt}[2]{\langle{#1}|{#2}\rangle}

\newcommand{\wv}[2]{\langle{#1}\rangle_{#2}}

\makeatletter
\title{Estimation of Spin-Spin Interaction by Weak Measurement Scheme}
\shorttitle{Estimation of Spin-Spin Interaction by Weak Measurement Scheme}
\author{Yutaka Shikano\inst{1,2} \and Shu Tanaka\inst{3}}
\shortauthor{Y. Shikano \and S. Tanaka}
\institute{                    
  \inst{1} Department of Physics, Tokyo Institute of Technology, Meguro, Tokyo, 152-8551, Japan \\
  \inst{2} Department of Mechanical Engineering, Massachusetts Institute of Technology, Cambridge, MA 02139, USA \\
  \inst{3} Research Center for Quantum Computing, Interdisciplinary Graduate School of Science and Engineering, \\
	Kinki University, Higashi-Osaka, Osaka, 577-8502, Japan
}
\abstract{
	Precisely knowing an interaction Hamiltonian is crucial to realize quantum information tasks, especially to experimentally demonstrate  
	a quantum computer and a quantum memory. We propose a scheme to experimentally evaluate the spin-spin interaction 
	for a two-qubit system by the {\it weak measurement} technique initiated by Yakir Aharonov and his colleagues. Furthermore, 
	we numerically confirm our proposed scheme in a specific system of a nitrogen 
	vacancy center in diamond. This means that the {\it weak measurement} can also be taken as a concrete example of the quantum process tomography.
}
\begin{document}
\maketitle
\section{Introduction}
Since the publication of the Shor algorithm~\cite{Shor}, that the prime factorization based on the current cryptography scheme can be solved with 
polynomial gates in quantum computation, many researchers have tried to realize a quantum computer in some physical systems~\cite{NC}. 
The realization of quantum computation is equivalent to designing the Hamiltonian adjusting to the given problem. 
Therefore, we have to precisely know a {\it specific} spin-spin interaction in a two-qubit system to realize quantum computation. We 
emphasize that the specific spin-spin interaction is quite different from an {\it effective} spin-spin interaction derived from 
the collective spins like the mean-field approximation. From the viewpoint of realizing quantum information processing tasks, 
knowing the effective Hamiltonian is futile since we cannot perfectly control a considered qubit.

Also, quantum storage devices, especially a quantum memory, is needed 
to realize quantum information processing tasks~\cite{DiVincenzo}. 
The vital properties of them are to be strong against decoherence, to write, and to read out easily. To construct them, 
we basically prepare two different qubit systems. One is easily manipulated and read out, and the other is insensitive
to decoherence, {\it i.e.}, its quantum information can be stored for a long time. 
Furthermore, we need to transfer quantum information for one qubit to the other, which is 
called to a quantum state transfer. There is the problem to exactly know the specific spin-spin interaction 
to perfectly realize the quantum state transfer. 
In a typical example of the Hamiltonian estimation, the dominant term of the Hamiltonian is assumed. 
Therefore, when we estimate the relative position between two spins from the measurement result, 
the Hamiltonian can be evaluated, {\it e.g.}, the ESR technique~\cite{ESR}. 
Since decoherence and an experimental error are inevitable in all physical systems on measuring the relative position between spins, 
it is difficult to perfectly transfer one qubit to the other for the considered two-qubit system. Therefore, we need an evaluation of the 
considered spin-spin interaction without knowing the relative position to improve an accuracy of quantum state transfer.

A quantum memory with small storage has been experimentally demonstrated in the systems between the 
polarization of a single photon and a single-electron spin in quantum dots~\cite{Kosaka} and between the polarization of a single 
photon and the $\Lambda$-type atom in the cavity quantum electrodynamics (cavity QED), which is called an electromagnetically 
induced transparency (EIT)~\cite{Honda}. The polarization of photons is used as the controlled device of quantum communication 
and quantum computation. The electron spin in quantum dots and the $\Lambda$-type atom in the cavity QED are 
used as the quantum storage devices. Also, a nitrogen vacancy (NV) center in diamond is a utmost potential example of the 
quantum storage devices. A NV electron spin state can be mapped to a nearby $^{13}$C nuclear spin state~\cite{Dutt}, 
which has a long coherence time at room temperature. Recently, a NV electron spin state has been also mapped to another one, 
which was located near the considered NV center, to improve the scalability of this system~\cite{Neumann}.

In this Letter, we propose a scheme to experimentally evaluate an unknown specific spin-spin interaction 
in a two-qubit system. We numerically confirm our proposal using an example 
of the NV center in diamond.

Our ultimate goal is precise parameter estimation of the specific spin-spin interaction, which is essential 
to realize quantum information processing tasks, especially a quantum memory, as discussed above. 
This is similar to the motivation of ref.~\cite{Maruyama} while they discussed this under different 
regulations. The key point of our proposal is to use the {\it weak measurement}~\cite{AAV} introduced as follows
by the different viewpoint to the original scheme.

By taking the {\it weak measurement} and the post-selection for a target system, 
we can obtain the {\it weak value} of an observable $A$ defined as $\wv{A}{w} := \bra{f}A\ket{i} / \bkt{f}{i}$,
where $\ket{i}$ and $\bra{f}$ are called a pre- and post-selected state, respectively~\cite{AAV}. 
These concepts are summarized in refs.~\cite{AV, shikano10}. 
This tool is useful to deeply understand quantum foundations, {\it e.g.}, to resolve the Hardy paradox~\cite{Aharonov} and 
to give an outlook on the macroscopic realism using the violation of the Leggett-Garg inequality~\cite{LG}. Therefore, many theoretical 
proposals to measure the weak value have recently proposed in some physical systems, {\it e.g.}, ref.~\cite{Romito}.
However, the problem of the weak value and the {\it weak measurement} is a little practical advantage, {\it e.g.}, 
the first demonstration of the spin Hall effect of light~\cite{HK} and the feedback control to stabilize the Sagnac interferometer~\cite{Dixon}. 
While the interaction between the target and probe systems is given in the conventional concept of the {\it weak measurement}, 
our addressed problem is that the interaction Hamiltonian is unknown and has to be evaluated~\footnote{Our proposed scheme is different 
from the {\it weak measurement tomography}; the post-selected state can be evaluated from the pre-selected state and the measured 
weak value~\cite{Shpitalnik}.}. 
In our proposal, the short-time evolution can be guaranteed in the {\it weak measurement} regime. 
In this Letter, we also show that the {\it weak measurement} 
can also be taken as a concrete example of the quantum process tomography. 
Therefore, our proposal gives a practical outlook on the {\it weak measurement} and the weak value.
\section{Proposed Protocol by Weak Measurement}
\begin{table*}[t]
\begin{center}
\begin{tabular}{|c|c|c|c|c|}
	\hline
	\multirow{2}{*}{$H_{int}$} & \multicolumn{4}{|c|}{Parameter Sets}   
	\\ \cline{2-5}
	& $\mathbf{r}_i$ & $\mathbf{p}$ & $\mathbf{q}$ & $\delta t [\mu {\textrm s}]$ \\ \hline
	\multirow{5}{*}{ $ \left( \begin{array}{ccc} 5 & -6.3 & -2.9 \\ -6.3 & 4.2 & -2.3 \\ -2.9 & -2.3 & 8.2 \end{array} \right)$} & $(0,0,1)$ 
	& $(0,0.59,0.81)$ &  $(-0.16,0,0.99)$ & $0.091$ \\ \cline{2-5}
	& $(-0.48,0.59,0.65)$ & $(0,0,1)$ & $(-0.25,0.59,-0.77)$ & $0.086$ \\ \cline{2-5}
	& $(-0.81,0.59,0)$ & $(-0.65,0.59,-0.48)$ & $(0.25,0.59,-0.77)$ & $0.073$ \\ \cline{2-5}
	& $(0,0,1)$ & $(0,0,1)$ & $(-0.99,0,-0.16)$ & $0.069$ \\ \cline{2-5}
	& $(0.81,0,-0.59)$ & $(-0.10,0.95,0.29)$ & $(0,0.81,-0.59)$ & $0.066$ \\ \cline{2-5}
	\multicolumn{1}{|r|}{(in MHz)} & $(0.31,0.95,0)$ & $(-0.18,0.95,-0.25)$ & $(0,0.81,0.59)$ & $0.051$ \\ \hline
\end{tabular}
\caption{Example of our proposed scheme. We apply our proposal to the NV center in diamond. The target and probe spins are the 
NV electron and nearby $^{13}$C nuclear spins, respectively. Also, the hyperfine interaction is given by eq. (15) in Supporting material in ref.~\cite{childress} 
but is taken as the unknown interaction to be evaluated. The set of the controlled parameters is decided from Fig.~\ref{dynamics}. It should be noted that 
the transposition is omitted in the columns of $\mathbf{r}_i, \mathbf{p},$ and $\mathbf{q}$ due to the limited page while they should be described as the row vector.}
\label{table1}
\end{center}
\end{table*}
Throughout this paper, we assume the followings:
\begin{enumerate}
	\item We can prepare a two-qubit system. One spin is called a target spin and the other is called a probe spin. 
		The total Hamiltonian is given by 
		\begin{equation}
			H_{tot} = H_t + H_p + H_{int}
		\end{equation}
		to simplify the discussion. Here, $H_t, H_p$, and $H_{int}$ denote the target spin Hamiltonian, the probe spin Hamiltonian, 
		and the interaction Hamiltonian between the target and probe spins, respectively.
	\item We can know the single spin dynamics: $H_t$ and $H_p$.
	\item We can manipulate and detect the state of the target and probe spins. Note that, the direct measurement for spins is not necessary. 
		After transferring the quantum state to another spin, measuring the transferred spin can be taken as the detection of the spin state 
		as in ref.~\cite{Dutt}.
\end{enumerate}
The above assumptions are realizable for some physical systems in recent quantum information technology.

Let us consider an {\it unknown} interaction Hamiltonian: 
\begin{align} 
	& H_{int} = \sum_{\mu, \nu \in \{ x, y, z \}}g_{\mu \nu} (\sigma^{\mu}_t \otimes \sigma^{\nu}_p), \notag \\
	& [ g_{\mu \nu} ] = \left( \begin{array}{ccc} g_{xx} & g_{xy} & g_{xz} \\ 
	g_{yx} & g_{yy} & g_{yz} \\ g_{zx} & g_{zy} & g_{zz} \end{array} \right) =: \left( {\mathbf n}_x \ {\mathbf n}_y \ {\mathbf n}_z \right) , \label{int}
\end{align}
where $\sigma^{\mu}_t$ and $\sigma^{\nu}_p$ are the Pauli matrices $(\mu, \nu = x, y, z)$ for target and probe spins, respectively.
The above equation represents the spin-spin interaction between target and probe spins. Here, the unknown parameters $g_{\mu \nu}$ are assumed to be 
a symmetric tensor and have six degrees of freedom. In the following, we propose a scheme to evaluate the parameters $g_{\mu \nu}$ of eq. (\ref{int}):
\begin{description}
\item[Step 1:] We prepare the pre-selected state for the target and probe spins denoted as  
	$\Phi_1 = \rho_t \otimes \rho_p$, 
	where $\rho_t = (I + {\mathbf r}_i \cdot {\mathbf \sigma}_t)/2$ and $\rho_p = (I + {\mathbf p} \cdot {\mathbf \sigma}_p)/2$ in the 
	Bloch sphere representation. Here, $I$ expresses the identity operator. ${\mathbf \sigma}_t$ and ${\mathbf \sigma}_p$ are the Pauli vectors.
	The controllable vectors, ${\mathbf r}_i$ and ${\mathbf p}$, are the Bloch vector or the Stokes vector for the 
	initial target spin and the initial probe spin, respectively.
\item[Step 2:] We wait for the short time $\delta t$. This procedure can be taken as the weak interaction regime by the {\it unknown} spin-spin 
	interaction. The state $\Phi_2 = e^{-i H_{tot} \delta t} \Phi_1 e^{i H_{tot} \delta t}$ is weakly entangled between target and probe spins. 
\item[Step 3:] We decide the post-selected state for the target spin ${\tilde{\rho}}_t = \Tr_p \Phi_2$ 
	by undertaking the quantum state tomography. 
	It should be noted that the quantum state tomography is not completed in a single event. We can repeat this procedure 
	since the state $\Phi_2$ is unchanged on remaining the conditions by Step 2. In principle, we perfectly undertake the quantum state 
	tomography for the target spin. 
\item[Step 4:] We measure the probe spin in the measurement direction $\tilde{{\mathbf q}}$ to obtain the expectation value as 
	\begin{equation} \label{real}
		\Ex ({\mathbf q} \cdot {\mathbf \sigma}_p) = \Tr (P({\tilde{{\mathbf q}}}) {\tilde{\rho}}_p),
	\end{equation}
	where ${\tilde{\rho}}_p := \Tr_t \Phi_2$, $P({\tilde{{\mathbf q}}}) := {\tilde{{\mathbf q}}} \cdot {\mathbf \sigma}_p$ is 
	the projection operator, and 
	${\mathbf q}$ is defined in the following step.
\item[Step 5:] We remove the contributions from the single spin Hamiltonians $H_t$ and $H_p$ by calculating the reverse 
	operation for the interaction time $\delta t$ 
	to the post-selected state ${\mathbf r}_f$ for the target spin and the modified measurement direction ${\mathbf q}$ for 
	the probe spin. Because of the short 
	interaction time $\delta t$, the parameters ${\mathbf r}_f$ and ${\mathbf q}$ can be defined as 
	$e^{i H_t \delta t} {\tilde{\rho}}_t e^{- i H_t \delta t} =: (I + {\mathbf r}_f \cdot {\mathbf \sigma}_t)/2$ and 
	$e^{i H_p \delta t} P(\tilde{{\mathbf q}}) e^{- i H_p \delta t} =: P ({\mathbf q})$ 
	by the Trotter-Suzuki decomposition~\cite{ST}.
\item[Step 6:] Changing the controllable parameters; the initial target spin state ${\mathbf r}_i$, the initial probe spin state ${\mathbf p}$, 
	the interaction time $\delta t$, and the measurement direction $\tilde{{\mathbf q}}$, we repeat the procedures from Step 1 to Step 5. 
\item[Step 7:] Since Steps 2 and 3 can be taken as the {\it weak measurement} regime, we obtain the expectation value in Step 4 under the first 
	order approximation as    
	\begin{align} \label{estimation}
		& \Ex ({\mathbf q} \cdot {\mathbf \sigma}_p) \approx {\mathbf q} \cdot {\mathbf p} \notag \\ 
		& + \sum_{\mu = x, y, z} \left[ 
		2 \, \delta t \frac{ \{ ({\mathbf q} \times {\mathbf n}_\mu ) \cdot {\mathbf p} \} 
		({\mathbf r}_i + {\mathbf r}_f) \cdot {\mathbf e}_\mu } {1 + {\mathbf r}_i \cdot {\mathbf r}_f} \right. \notag \\
		& \left. \ \, + 2 \, \delta t \frac{ [ {\mathbf n}_\mu \cdot {\mathbf q} - ({\mathbf n}_\mu \cdot {\mathbf p})
		({\mathbf q} \cdot {\mathbf p}) ] ( {\mathbf r}_i \times {\mathbf r}_f ) \cdot {\mathbf e}_\mu }{1 + 
		{\mathbf r}_i \cdot {\mathbf r}_f} \right], 
	\end{align}
	where ${\mathbf e}_\mu$ is a unit vector in the $\mu$ direction ($\mu = x, y, z$).
	Note that, the above equation includes the weak value of ${\mathbf \sigma}_s$,   
	$\wv{ {\mathbf \sigma}_s}{w} = [{\mathbf r}_i + {\mathbf r}_f + i ({\mathbf r}_i 
	\times {\mathbf r}_f)]/(1 + {\mathbf r}_i \cdot {\mathbf r}_f)$ 
	and is derived by the straightforward extension of ref.~\cite{Wu}. 
	We obtain the six linear equations by inserting the controlled parameters in Step 6. 
	Then, we can decide the six unknown parameters $[g_{\mu \nu}]$.
\end{description}
The six linear equations can be expressed by $\mathbf{\xi} = A^{-1} \mathbf{\zeta}$, where
\begin{align}
	{\mathbf \xi} & := (g_{xx}, g_{yy}, g_{zz}, g_{xy}, g_{xz}, g_{yz})^{T}, \label{estimation1} \\
	{\mathbf \zeta} & := \{ \Ex ({\mathbf q}_k \cdot {\mathbf \sigma}_p) - {\mathbf q}_k \cdot {\mathbf p}_k \} 
	\frac{1 + {\mathbf r}_{i,k} \cdot {\mathbf r}_{f,k}}{2 \, \delta t_{k}}, \label{weakcondition} \\
	[A]_{j,k} & := ({\mathbf p}_k \times {\mathbf q}_k)_\mu ({\mathbf r}_{i,k} + {\mathbf r}_{f,k})_\nu \notag \\
	& \ \ \ \ \  + \{ {\mathbf q}_k - {\mathbf p}_k
	({\mathbf q}_k \cdot {\mathbf p}_k) \}_\mu ({\mathbf r}_{i,k} \times {\mathbf r}_{f,k})_\nu. \label{estimation2} 
\end{align}
Here, the suffix $k$ denotes the subscript labeling on the experimental data and $T$ denotes the matrix transpose. 
$ \Omega := \{ (\mu, \nu) \} = \{ (x,x), (y,y), (z,z), (x,y), (x,z), (y,z) \}$ corresponds to the suffix $\{ j \} = \{ 1, 2, 3, 4, 5, 6 \}$  
since $[g_{\mu \nu}]$ is the symmetric tensor. Therefore, $A$ is the square matrix of order six.
\begin{figure}[bt]
\begin{center}
	\includegraphics[width=8cm]{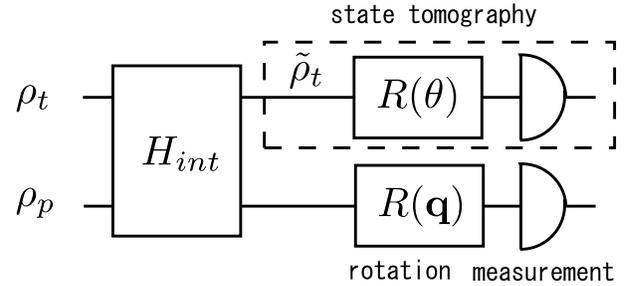}
	\caption{Quantum circuit representation of our proposal. $R (\theta)$ and $R ({\mathbf q})$ express the rotation operation of the spin for 
	an arbitrary angle $\theta$ and the measurement direction ${\mathbf q}$ for some directions, respectively. The dashed-line area indicates 
	the procedures of the quantum state tomography by changing the angle $\theta$.}
	\label{fig1}
\end{center}
\end{figure}
Our proposal is summarized in Fig.~\ref{fig1} when the single spin Hamiltonians, $H_t$ and $H_p$, are not considered because of Step 5. 
Compared to the established schemes of the quantum process tomography, this advantage is to evaluate the specific spin-spin interaction 
without knowing the relative position between spins~\cite{childress}, using the long-time dynamics~\cite{Devitt}, 
and using the Bell measurement~\cite{Mohseni}. Furthermore, we would like to emphasize that the operations to the target and 
probe systems are different due to the linearity of the {\it weak measurement} while we have to use the same 
operations to the target and probe systems from the viewpoint of the conventional quantum process tomography.
\begin{figure}[htb]
\begin{center}
\includegraphics[width=8cm]{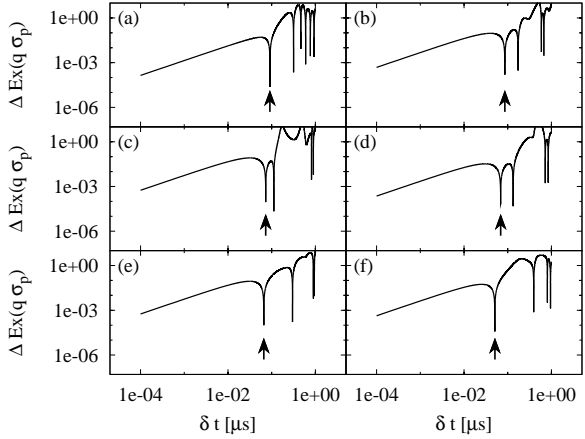}
\end{center}
\caption{Dynamics of the probe spin. Under the controlled parameters of Table~\ref{table1}, in which (a)--(f) correspond sequentially from the top, the dynamics 
of the expectation values of the probe spin is numerically shown. The vertical axis is $\Delta \Ex ({\mathbf q} \cdot {\mathbf \sigma}_p)$, which is defined as the 
absolute value of the differences between the right-hand side of Eqs.~(\ref{real}, \ref{estimation}), {\it i.e.}, the second- and higher-order correction, 
and the horizontal axis is the time scale $\delta t$ 
of the order of micro seconds. The time at which $\Delta \Ex ({\mathbf q} \cdot {\mathbf \sigma}_p)$ is small has to be chosen at marked positions. 
Since we have not analytically obtained the condition to appear in the dent of $\Delta \Ex ({\mathbf q} \cdot {\mathbf \sigma}_p)$, 
we now search the points by hand. In practical, these points have to be picked up from many trials changing the set of the controlled parameters. 
Since the interaction Hamiltonian is uniquely determined, we can find out these by the above process.}
\label{dynamics}
\end{figure}
\section{Example in Nitrogen Vacancy Center in Diamond}
In the final part of this Letter, as an example to show the reliability of our proposed scheme, 
we apply this to evaluate the hyperfine interaction in a system of a NV center in diamond~\cite{childress} under the above assumption~\footnote{Since a 
position of the $^{13}$C nuclear spin cannot be directly observed in current technology, our proposal seems to be essentially needed.}. 
Here, the target and probe spins are the NV electron and nearby $^{13}$C nuclear spins, respectively.
Because of Step 5, the single spin Hamiltonians can be ignored; we assume $H_t = H_p = 0$ in the numerics, {\it i.e.}, $\tilde{{\mathbf q}} = {\mathbf q}$.
Under a given interaction Hamiltonian, we numerically calculate the time evolution from the given initial state. 
Then, we calculate the final state for the target spin and the expectation value of the probe spin. By Eqs. (\ref{estimation1}) -- (\ref{estimation2}), 
we numerically evaluate the interaction Hamiltonian. We compare the given and evaluated interaction Hamiltonians by 
the error defined as the mean value and the standard deviation between the given Hamiltonian $[g_{\mu \nu}]$ and 
the evaluated one $[{\tilde{g}}_{\mu \nu}]$:
${\rm error} := \wv{g}{} \pm \sigma_g$, where 
\begin{align}
\wv{g}{} & := \sum_{(\mu, \nu) \in \Omega } ( {\tilde{g}}_{\mu \nu} - g_{\mu \nu} ) / 6, \\ 
\sigma_g & := \sqrt{ \sum_{(\mu, \nu) \in \Omega } \{ ({\tilde{g}}_{\mu \nu} - g_{\mu \nu}) - \wv{g}{} \}^2 / 5}.
\end{align}
From the example of Table~\ref{table1}, we obtain the evaluated Hamiltonian, 
\begin{equation}
	\left( \begin{array}{ccc} 4.98 & -6.29 & -2.92 \\ -6.29 & 4.21 & -2.30 \\ -2.92 & -2.30 & 8.35 \end{array} \right).
\end{equation}
It is noted that how to decide the choice of the estimated parameters is not mentioned in this Letter. 
However, they can be in principle picked up from many trials. 
Our proposed scheme can reproduce the original hyperfine interaction with high precision 
since the evaluated error, $0.022 \pm 0.063 \ {\rm MHz}$, is within $1 \%$. It is noted that the parameters, $\mathbf{r}_i, \mathbf{p},$ and $\mathbf{q}$, 
are determined from Fig.~\ref{dynamics}. 
Therefore, our proposal may be experimentally realizable to evaluate the spin-spin interaction.
\section{Concluding Remark and Outlook}
In conclusion, we have proposed the procedures from Step 1 to Step 7 to experimentally evaluate the specific spin-spin interaction (\ref{int}) 
in a considered two-qubit system from eq. (\ref{estimation}) obtained in the {\it weak measurement} 
technique. This gives the practical advantage to the {\it weak measurement} / weak value analysis. The {\it weak measurement} 
can be also taken as an example of the quantum process tomography and quantum estimation.

There are remaining problems on our proposal. First, we have not adjusted our proposal under realistic experimental constraints. 
As the first step, we roughly considered the realistic experimental setup in the NV center in diamond~\cite{SKTH}.
Second, we have not yet given an optimized method to choose the controllable parameters while  
the abstract estimation limit is shown~\cite{holger}, which is related to the channel-parameter 
estimation problem~\cite{Fujiwara}. 
Third, we have not mentioned the scalability~\cite{DiVincenzo} of our proposal. 
While it may be possible to overcome the scalability under the assumption 
of the spin configuration like ref.~\cite{Maruyama}, there are many varieties on this problem by the considerable physical qubit.
Finally, we did not consider the decoherence effects during this scheme. According to ref.~\cite{shikano10}, 
it can be included in the decoherence effect for the weak value. 
However, this can be ignored since our proposal only uses the short-time interaction. This might be a practical 
advantage of our proposal compared with the established schemes. We believe that {\it weak measurement} technique will 
be a powerful tool to analyze the fundamental properties in condensed matter systems 
and quantum optics like our study, despite its naming.
\acknowledgments
The authors would like to acknowledge useful discussions 
with Yakir Aharonov, Daniel Burgarth, Akio Hosoya, Hideo Kosaka, Koji Maruyama, Makoto Negoro, James Rabeau, 
Shuhei Tamate, Norman Yao and Tomohiro Yoshino, and also
thank the Supercomputer Center, Institute for Solid State Physics, 
University of Tokyo for use of the facilities.
This work is supported by JSPS Research Fellowships for Young Scientists (No. 21008624), 
Grand-in-Aid for Young scientists Start-up (No. 21840021), 
and Grand-in-Aid for Scientific Research (B) (No. 22340111) from JSPS, 
and ``Open Research Center" Project for Private Universities.

\end{document}